\documentclass[prl,aps,showpacs,twocolumn,superscriptaddress]{revtex4}
\usepackage{amsmath,amsfonts,amssymb,graphics,graphicx,epsfig,color,times}
\usepackage[utf8]{inputenc}
\usepackage{cancel,ifthen,ulem}

\DeclareMathSymbol{\epsilon}{\mathord}{letters}{"22}
\DeclareMathSymbol{\theta}{\mathord}{letters}{"23}
\DeclareMathSymbol{\rho}{\mathord}{letters}{"25}
\DeclareMathSymbol{\phi}{\mathord}{letters}{"27}
\DeclareMathSymbol{\varepsilon}{\mathord}{letters}{"0F}
\DeclareMathSymbol{\vartheta}{\mathord}{letters}{"12}
\DeclareMathSymbol{\varphi}{\mathord}{letters}{"1E}
\DeclareMathSymbol{\varrho}{\mathord}{letters}{"1A}

\renewcommand{\vec}[1]{\text{\boldmath$#1$}}

\newcommand{\hr}{\hat{\rho}}
\newcommand{\hdo}{\hat{\kappa}}
\newcommand{\Tr}{\mathrm{Tr}}
\newcommand{\hsig}{\hat{\sigma}}

\begin{document}

\title{Dynamical simulation of integrable and non-integrable models in the Heisenberg picture}
\author{Dominik Muth}
\email{muth@physik.uni-kl.de}
\affiliation{Fachbereich Physik und Forschungszentrum OPTIMAS, Technische Universit\"at
Kaiserslautern, D-67663 Kaiserslautern, Germany}
\affiliation{Graduate School Materials Science in Mainz, Technische Universit\"at
Kaiserslautern, D-67663 Kaiserslautern, Germany}
\author{Razmik G. Unanyan}
\affiliation{Fachbereich Physik und Forschungszentrum OPTIMAS, Technische Universit\"at
Kaiserslautern, D-67663 Kaiserslautern, Germany}
\author{Michael Fleischhauer}
\affiliation{Fachbereich Physik und Forschungszentrum OPTIMAS, Technische Universit\"at
Kaiserslautern, D-67663 Kaiserslautern, Germany}

\begin{abstract}
The numerical simulation of quantum many-body dynamics is typically limited by the linear growth of entanglement with time. Recently numerical studies have shown, however, that for 1D Bethe-integrable models the simulation of local operators in the Heisenberg picture can be efficient as the corresponding operator-space entanglement grows only logarithmically. Using the spin-1/2 XX chain as generic example of an integrable model that can be mapped to free fermions, we here provide a simple explanation for this. We show furthermore that the same reduction of complexity applies to operators that have a high-temperature auto correlation function which decays slower than exponential, i.e., with a power law. This is amongst others the case for models where the Blombergen-De Gennes conjecture of high-temperature diffusive dynamics holds. Thus efficient simulability may already be implied by a single conservation law (like that of total magnetization), as we will illustrate numerically for the spin-1 XXZ model.
\end{abstract}


\date{\today}
\maketitle


\renewcommand{\vec}[1]{\text{\boldmath$#1$}}





White's Density Matrix Renormalization Group (DMRG) \cite{White1992} and
it's more recent generalizations to time evolution using the Time Evolving Block Decimation (TEBD) \cite{VidalCombined} or t-DMRG \cite{DaleyWhite} algorithms are
indispensable tools in the numerical simulation of one-dimensional quantum many body 
systems. They permit high-accuracy calculations, provided that
the entanglement between any two complementary partitions remains small.
For finite-range interactions this is the case for the ground state \cite{Verstraete2006}. 
However in real time evolution the entanglement 
often grows linear in time, limited only by the Lieb-Robinson upper bound
\cite{Lieb1972,BE}. E.g. for the spin-$\frac12$ XY chain the evolution of the 
entanglement entropy was investigated  in  \cite{Fagotti2008} showing explicitly the linear growth
in time. 

However the evolved state contains a lot of information which is of little
interest. Experimental measurements as well as theories are almost solely
concerned with few particle properties, i.e., quantities that can be
expressed in terms of expectation values of only a small number of
elementary operators. This suggests to go to the
Heisenberg picture (HP) instead, and to simulate the dynamics of these operators.
Prosen et al. \cite{Prosen2007a,Prosen2007} where the first to
pursue this approach. 
They observed an exponential speed up in
numerical simulations of local operators for integrable systems. 
So far there is however no general understanding of why this is the case and
whether or not integrability is crucial. In the present
paper we provide an explanation of the speed-up for integrable models
that can be mapped to free fermions.
We also argue that integrability is not necessary and that 
the existence of a conservation law may suffice
for the efficient simulation of the dynamics of local operators that 
constitute the conserved quantity. 
We will discuss the spin-$\frac{1}{2}$ and
spin-1 XXZ models as specific examples supporting and illustrating our arguments.

In order to do HP simulations using e.g. the TEBD scheme, the operator $\hat O(t)$
at time $t$ is expressed in terms of a matrix product operator (MPO). For typical 
observables this is straightforward for the initial time $t=0$.
Time evolution is then calculated by updating the matrices according to the Heisenberg 
equation of motion using a Trotter decomposition.
Efficient simulation requires that the matrix dimension of 
the MPO's (called bond dimension) is limited to a maximum value $\chi $. 
This means that only the $\chi $ largest Schmidt values \textit{in the Hilbert space
of operators} are kept, corresponding to a small operator-space (OS) entanglement between any two 
complementary partitions of the lattice. 
To quantify the entanglement of an operator $\hat O(t)$, which after
proper normalization can be viewed as state vector in OS, we
use the OS R\'{e}nyi entropies 
(OSRE):
\begin{equation}
\ S_{\alpha }=\frac{\log _{2}\mathrm{Tr}\ \hdo%
^{\alpha }}{1-\alpha }\ge S_\beta,\quad \beta > \alpha >0
\end{equation}
Here $\hdo$ is the corresponding reduced density matrix in OS resulting from tracing out
the left or right partition at a given bond. In the limit $\alpha \rightarrow 1$, $S_\alpha$ is the well known von
Neumann entropy, which is a good measure of bi-partite entanglement. For $\alpha \to 0$, $S_\alpha$  gives the 
dimension of the Hilbert space.
Clearly for an MPO of bond dimension $\chi$, the maximum for all R\'{e}nyi entropies
is $\log_2 \chi$. Although it is not yet fully established when a quantum state or an operator is faithfully represented by a matrix product
with finite bond dimension, one can employ the results of Schuch et al. \cite{Schuch2008a} 
to show that efficient simulation is impossible if the R\'{e}nyi entropies with $\alpha > 1$ 
scale faster than logarithmically with time. If $S_{\alpha>1 }$ grows linearly in time, we must expect
that the computational cost required to reach a certain accuracy, which is polynomial in $\chi $, 
will grow exponentially with time (note that this is not necessarily true for $S_\alpha$ with  
$\alpha \leq 1$ \cite{Schuch2008a}). 
In fact for the time evolution of typical state vectors
in the Schr\"odinger picture this is very often the case \cite{Schuch2008}.
On the other hand an at most logarithmic growth of $S_{\alpha>1}$ is 
a necessary condition for an efficient simulability. Although not sufficient, it also gives good indication when such a simulation is possible.
In the following we will discuss the time evolution of the 
OSRE for a generic model, the XXZ chain, $\hat{H} = -\frac{1}{2}%
\sum_j\left(\hat{\sigma}^x_j\hat{\sigma}^x_{j+1}+\hat{\sigma}^y_j%
\hat{\sigma}^y_{j+1}+\Delta\hat{\sigma}^z_j\hat{\sigma}^z_{j+1}\right)$,
where $\hat{\sigma}^{x,y,z}$ denote the Pauli matrices in the spin-$\frac12$
case and the spin-1 matrices (eigenvalues $-1,0,1$) in the spin-1 case
respectively. The spin-$\frac12$ case is integrable for any value of the
anisotropy $\Delta$. For the special case of $\Delta =0$ (spin-$\frac12$ XX model) this model
can be mapped to free fermions.

\paragraph{integrable models equivalent to free fermions:}
Let us consider the spin-$\frac12$ XXZ model as a generic
example of a 1D integrable model.  We have calculated the
time evolution of the OSRE $S_2$ for different types of 
simple operators using the TEBD scheme with open boundary
conditions and a fourth order Trotter decomposition \cite{Sornborger1999}. The restriction to open boundary conditions is not an issue for
local operators as long as the time is shorter than the propagation time  to reach
the boundaries \cite{Lieb1972}.
Although not shown the OS von-Neumann entropy $S_1$ has the
same scaling behavior.
One clearly notices that the OSRE of all operators scales at most logarithmically in time,
an observation made already by Prosen et al. for other integrable models \cite{Prosen2007,Pizorn2009a}.
In the special case of $\Delta =0$ the entropy even saturates at a finite value for some
operators like $\hat \sigma^z$ or products at a small number of different lattice sites. 

\begin{figure}[tbh]
\epsfig{file=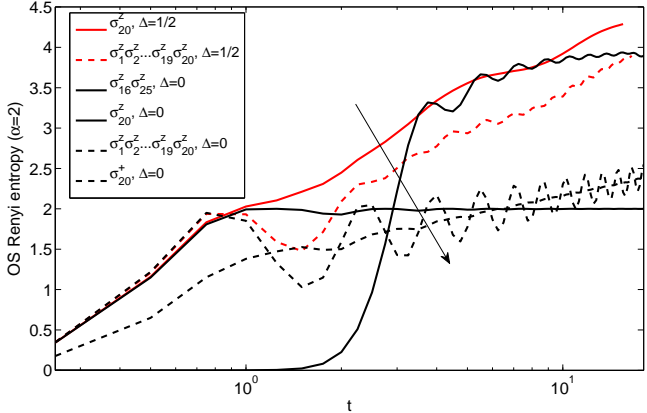,width=\columnwidth}
\caption{(Color online) OSRE dynamics for the 40 site spin-$\frac{1}{2}$ XXZ
model for a split in the center. The legend gives initial operator and
anisotropy in the order in which the arrow cuts the graphs. Dashed lines
mark infinite index operators (see text). $\protect\chi =1000$ is used in all cases and
the numerical error is negligible on the time scale shown.}
\label{fig:spin12}
\end{figure}

In the following we will provide an explanation of the entropy scaling for the case of the
XX-model, i.e., for $\Delta =0$, which can be mapped to free fermions. This will be done
by reexpressing the XXZ model in terms of  Majorana-fermion operators \cite{Prosen2008}, which turns out to be 
more convenient than the more common Wigner-Jordan transformation: $\hat{w}_{2j-1}=\Bigl( \prod_{l<j}\hat{\sigma}_{l}^{z}\Bigr) \hat{\sigma}_{j}^{x}$, $\hat{w}_{2j}=\Bigl( \prod_{l<j}%
\hat{\sigma}_{l}^{z}\Bigr) \hat{\sigma}_{j}^{y}$.
 The Majorana operators are Hermitian and
fulfill anti-commutation relations $\left\{ \hat{w}_{j},\hat{w}_{l}\right\}
=2\delta _{jl}$. The three types of interactions in the XXZ-model can be
reexpressed as 
\begin{eqnarray}
 \hat{\sigma}_{j}^{x}\hat{\sigma}_{j+1}^{x} &=& -i\hat{w}_{2j}\hat{w}_{2(j+1)-1},\nonumber\\ 
\hat{\sigma}_{j}^{y}\hat{\sigma}_{j+1}^{y} &=& i\hat{w}_{2j-1}\hat{w}_{2(j+1)},\\ \hat{\sigma}_{j}^{z}\hat{\sigma}_{j+1}^{z} &=& -\hat{w}_{2j-1}\hat{w}_{2j}
\hat{w}_{2(j+1)-1}\hat{w}_{2(j+1)}.\nonumber
\end{eqnarray}
A complete basis in the
OS is given by  $\hat{P}_{\alpha }=\prod_{j}\hat{w}%
_{2j-1}^{\alpha _{2j-1}}\hat{w}_{2j}^{\alpha _{2j}}$, where $\alpha \equiv(\alpha_1,\alpha_2,\dots)$ and  
$\{\alpha_l \}\in \{0,1\}^{N}$. We can now define adjoint-fermion annihilators and creators via $\hat{a}_{j}|\hat{%
P}_{\alpha }\rangle =\alpha _{j}|\hat{w}_{j}\hat{P}_{\alpha }\rangle $, $%
\hat{a}_{j}^{\dagger }|\hat{P}_{\alpha }\rangle =(1-\alpha _{j})|\hat{w}_{j}%
\hat{P}_{\alpha }\rangle $, with $\{ \hat{a}_{j},\hat{a}_{l}^{\dagger
}\} =\delta _{jl}$. Associating the adjoint vacuum $|\hat P_0\rangle$ with the unity operator
$\mathbf{1}$, i.e., $|\mathbf{1}\rangle =|\hat{P}_{0}\rangle $, we can express
all operators in terms of adjoint-fermion
excitations \cite{Pizorn2009a}: $|\hat{P}_{\alpha }\rangle =\prod_{j} \bigl(\hat{a}_{2j-1}^{\dagger }\bigr) 
^{\alpha_{2j-1}} \bigl(\hat{a}_{2j}^{\dagger }\bigr) ^{\alpha _{2j}} |\mathbf{1}\rangle $.
Mapping the Heisenberg equation then gives a Schr\"odinger like
equation for the evolution in OS,
\begin{equation}
i\frac{d}{dt}\hat{P}_\alpha = \left[ \hat{P}_\alpha, \hat{H}\right] \
\mapsto\ i\frac{d}{dt}\vert \hat{P}_\alpha \rangle = \left\vert \left[ \hat{P%
}_\alpha, \hat{H}\right] \right\rangle=: \hat{\cal H}\vert \hat{P}_\alpha
\rangle.
\end{equation}
with a ``super''-Hamiltonian $\hat{\cal H}$.
Explicitly calculating the terms in the commutator for the XX-model via
$\vert [ \hat{P}_\alpha, \hat{\sigma}^x_{j}\hat{\sigma}^x_{j+1}
] \rangle = 2i\bigl(\hat{a}^\dagger_{2j}\hat{a}_{2(j+1)-1}-\text{h.a.}\bigr)\vert \hat{P}_\alpha \rangle$,
and  $\vert [ \hat{P}_\alpha, \hat{\sigma}^y_{j}\hat{\sigma}^y_{j+1}] \rangle = -2i\bigl(\hat{a}^\dagger_{2j-1}\hat{a}_{2(j+1)}-\text{h.a.}\bigr)\vert \hat{P}_\alpha \rangle$
yields the XX super-Hamiltonian
\begin{eqnarray}
 \hat{\cal H}_{XX} = i\sum_j \left(\hat a_{2j}^\dagger \hat a_{2j+1} 
+\hat a_{2j-1}^\dagger\hat a_{2j+2} 
-\text{h.a.}\right)
\end{eqnarray}
This Hamiltonian corresponds to two uncoupled chains of free fermions.
The total number of adjoint fermions, $\sum_{m=1}^{2N}\hat{a}%
^\dagger_m\hat{a}_m$, is conserved.
Note that the anisotropy $\Delta$ in the original XXZ Hamiltonian would introduce recombination
and pair creation across the chains. Although the above 
mapping is non local, operators acting only left of a given site $j$
will be mapped to fermions that are again only left of this very site. So
the OSRE of the original XX-model will be the same as the corresponding
state vector R\'enyi entropy of two uncoupled chains of free fermions.
Thus we have to calculate the entanglement dynamics of the two uncoupled
chains with an initial state given by the operator in questions to get the
OSRE in the XX model. The key point is that local operators are equivalent to very
special, simple initial states in the corresponding fermion chains.
We here have to distinguish between
finite index operators (those that involve only a finite number of
adjoint fermions after the mapping) and infinite index operators (involving a number proportional to the system size $L$). An example of the first
kind is $|\hat{\sigma}_{j}^{z}\rangle =-i\hat{a}
_{2j-1}^{\dagger }\hat{a}_{2j}^{\dagger }|\mathbf{1}\rangle $. Examples of the
second kind arise either from local operators
like $|\hat{\sigma}_{j}^{x}\rangle =i^{j-1}\left( \prod_{l=1}^{2(j-1)}\hat{a}%
_{l}^{\dagger }\right) \hat{a}_{2j-1}^{\dagger }|\mathbf{1}\rangle $ or non
local ones like $|\hat{F}\rangle =\left\vert \prod_{l=1}^{j-1}\hat{\sigma}%
_{l}^{z}\right\rangle =i^{j-1}\left( \prod_{l=1}^{2(j-1)}\hat{a}%
_{l}^{\dagger }\right) |\mathbf{1}\rangle $. 

We proceed by showing that the bi-partite R\'enyi entropy $S_2$ for a system of free fermions in 1D 
is strictly related to the number fluctuations in any one of the two partitions
assuming a fixed total number. We can assume that the initial state of the fermions corresponding to the local
operators of interest is a
Gaussian state. Due to the free evolution it remains Gaussian and can be transformed into a product form $\hr = \bigotimes_j\hr_j$ where the $\hr_j$ correspond to site $j$ and have eigenvalues $\frac{1\pm\eta_j}{2}$, $\vert\eta_j\vert\le1$. The square of the variance of the total particle number in each partition
is then $\Delta N^2_A=\sum_{j\in A}(1-\eta_j^2)/{4} =\Delta N^2_B$ \cite{Klich2006}. On the other hand $S_2=-\log_2\Tr{\hr^2} = -\sum_j \log_2\left(1-(1-\eta_j^2)/{2}\right)$. Using $\frac{2}{\ln2}\frac{x}{2-x}\le-\log_2\left(1-x\right)\le\frac{1}{\ln2}\frac{x}{1-x}$, where $0\le x \le \frac12$, one obtains
\begin{equation}
\frac{4}{\ln 2}\Delta N^{2} \geq S_{2} \geq \frac{2}{\ln 2}\Delta N^{2}.
\end{equation}

For finite index operators we find saturation as can be seen in Fig. \ref{fig:spin12}. This reflects the fact, that
there is only a finite number $M$ of free particles present in both chains together. Thus a
finite $\chi$ of $2^M$ yields the exact solution \cite{HC} for all times \footnote{$S_{2}$ remains finite also because $\Delta N^2 \xrightarrow{t\rightarrow\infty}\frac{M}{4}$ as the probability for a particle of being left or right becomes equal.}.
For infinite index operators we observe logarithmic growth of the OSRE,
see Fig. \ref{fig:spin12}. While the infinite number of involved
adjoint fermions may suggest a linear growth of $\Delta N^2$, this is not the case as
can be understood in the following way:
The super state corresponding to a infinite index operator like  $|\hat{F}\rangle $ (a finite size
example of which is shown in Fig. \ref{fig:spin12}) is filled up completely
with fermions in the left part of the chains. Inside these regions the Pauli
principle prevents hopping of fermions and thus only particles
at the edge where the effective band-insulator is connected to the vacuum 
can move and fill the empty parts of the double chain.
For the half filled chain Antal et al. have shown that $\Delta N^{2}\cong (\ln t+D)/2\pi^2$ in the limit of large $t$ with a known constant $D>0$ \cite{Antal2008}.
Other infinite index operators that result in a initial
occupation of the two chains different from that of $\vert \hat{F} \rangle$
only on a finite number of sites show the same logarithmic long time behavior
of the OSRE, see $%
\vert \hat{\sigma}^+ \rangle$ at a single site, also shown in Fig. \ref{fig:spin12}.
This explains the dynamics of the OSRE in the XX model as a
generic example of an integrable model that can be mapped
to free fermions.

\paragraph{non-integrable models:}
We now show that there is another class of systems and operators which
may allow an efficient simulation of dynamics in the HP.
We construct an upper bound for the OSRE $S_{\alpha }$, $\alpha
>1$, in terms of the infinite-temperature auto-correlation function
(ITAC). Without loss of generality we assume 
a normalized operator, i.e. $\frac{1}{d^L}\mathrm{Tr}\left[ {\hat{O}}^{\dagger }\hat{O}\right] =1$, where $d$ is the local dimension of the chain. With
respect to a splitting of the chain of length $L$ into two parts here and below all $\hat{A}$ act on
the sub chain A of length $L_{\rm{A}}$ and all $\hat{B}$ on B of length $%
L_{\rm{B}}$. 
Any operator can be represented as
$
\hat{O}\left( t\right) = \sum_{m,n}\Lambda_{mn}\left( t\right) \hat{A}_{m}\otimes \hat{B}_{n}
$
with orthonormal bases $\frac{1}{d^{L_{\rm A}}}\mathrm{Tr}\left[ \hat{A}_{n}{}^{\dagger }\hat{A}%
_{m}\right] = \frac{1}{d^{L_{\rm B}}}\mathrm{Tr}\left[ \hat{B}%
_{n}^{\dagger }\hat{B}_{m}\right] =\delta _{nm}$.
$\Lambda $\ is a matrix and its singular values $\sqrt{\lambda
_{n}},$ ($\lambda _{1}\geq \lambda _{2}\geq\dots$ are the eigenvalues of $\hdo$)
are coefficients of a Schmidt decomposition $\hat{O}(t)=\sum_{n=1}^\chi
\sqrt{\lambda _{n}}\hat{\cal A}_{n}(t)\otimes \hat{\cal B}_{n}(t)$, where now the Schmidt rank $\chi$ is at most $d^{2\min(L_{\rm A}, L_{\rm B})}$. 
This allows to express the infinite-temperature auto-correlation function
in terms of Schmidt coefficients. 
We find for $\alpha > 1$
\begin{eqnarray}
\left\vert \left\langle \hat{O}^{\dagger }(t)\hat{O}\right\rangle _{T=\infty
}\right\vert
&=&\left\vert \mathrm{Tr}\left[ \Lambda ^{\dagger }\left( t\right) \Lambda
\left( 0\right) \right] \right\vert \nonumber\\
&\leq& \sum_{k=1}^{\chi}\sqrt{\lambda
_{k} \lambda _{k}\left( 0\right) }
\label{eq:vN}
\end{eqnarray}
\begin{eqnarray}
&\le& \Tr\sqrt{\hdo(0)} \left(\sum_{k=1}^{\chi}\frac{\sqrt{\lambda_k(0)}}{\Tr\sqrt{\hdo(0)}}\lambda_k^{\frac{\alpha}{2}}\right)^{\frac1\alpha}
\label{eq:J}\\
&=&\left(\Tr\sqrt{\hdo(0)}\right)^{1-\frac1\alpha} \left(\sum_{k=1}^{\chi}\lambda_k^{\alpha}\right)^{\frac{1}{2\alpha}}\label{eq:CS}
\end{eqnarray}
In (\ref{eq:vN}) we made use of von Neumann's trace inequality (see e.g. \cite{Gantmacher1959}). Furthermore Jensen's inequality can be used because $x^{\frac1\alpha}$ is a concave function in $x$. Finally (\ref{eq:CS}) is true by the  Cauchy-Schwarz inequality.
We thus obtain the following estimate for R\'{e}nyi entropies, assuming an initial product operator, $\Tr\sqrt{\hdo(0)}=1$, for simplicity:
\begin{equation}
S_{\alpha }\leq \frac{2\alpha }{1-\alpha }\log _{2}\left\vert \left\langle
\hat{O}^{\dagger }(t)\hat{O}\right\rangle _{T=\infty }\right\vert\quad \text{\
for }\alpha >1.  \label{upperBound}
\end{equation}
If the ITAC decays with a power law or slower in time, $S_{\alpha }$ will
grow at most logarithmically for $\alpha >1$. The ITAC has been studied over
decades in condensed matter physics as it is measured in nuclear magnetic
resonance and neutron scattering experiments in magnetic spin chains. While
not proofed rigorously, it is believed that the Blombergen-de Gennes
conjecture \cite{BGK} of spin diffusion holds: If $%
\sum_{j=1}^{L}\hat{O}_{j}$ is a conserved quantity, then the ITAC of $%
\hat{O}_{j}$ will show diffusive behavior (i.e. $\sim 1/\sqrt{t}$ in 1D). To
our knowledge there is no counter example except for integrable models,
where this diffusive behavior can turn into a ballistic one (i.e. $\sim 1/t$
in 1D) \cite{Fabricius1998,Sirker2006}. Nevertheless it always remains slower than
exponential. We conclude that in the HP TEBD we can expect $S_{2}$ to grow at
most logarithmically in time, even if the model is non-integrable, if the initial operator belongs to a
conservation law (for integrable systems there is an infinite number of
those, but one is sufficient). This in turn indicates that an efficient
classical simulation should be possible for large times.

\begin{figure*}
\epsfig{file=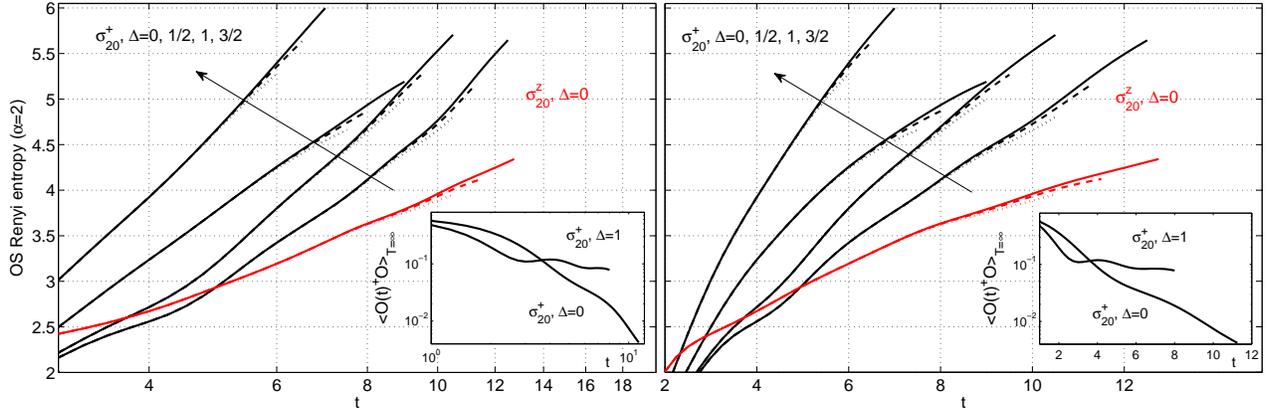,width=.94\textwidth}
\caption{(Color online) OSRE dynamics for the 40 site spin-1 XXZ model for a
split in the center. Dotted, dashed, and solid lines indicate
simulations using $\protect\chi =300$, $\protect\chi =500$, and $\protect%
\chi =1000$ respectively. The left panel features a logarithmic, the right panel 
a linear time scale. The curves show clear indication of the predicted long-time scaling. The insets show two corresponding ITAC curves for $\hsig^+_{20}$ (note the logarithmic vertical scaling).
\label{fig:spin1}
}
\end{figure*}

The spin-1 XXZ chain is an example of a non-integrable system, although extension to
additional higher-order nonlinear terms may turn it into an integrable one \cite{Babujian1986,Sogo1984}.
However the
total z-magnetization $\sum_{j=1}^L \hat{\sigma}^z_j$ is conserved. 
This conservation law will lead to a logarithmic scaling of $S_2$ for $\hat \sigma^z$. Fig. \ref{fig:spin1} shows numerical indication for this. It should be noted that the spin-1 model is
computationally much harder than the spin-$\frac12$ model since the local Hilbert
space dimension is increased. Although we do observe logarithmic scaling of the OSRE
corresponding to $\hat \sigma^z$, the
prefactor is large, such that we can not go to far in time. The plot shows data for different
matrix dimension $\chi$ up to the point where the cutoff error becomes
substantial. A clear tendency is visible: On the logarithmic scale $%
S_2$ approaches a straight line,
while in the linear plot a sub-linear
scaling is evident.  This is consistent with the expected
logarithmic scaling of the OSRE.
For $\hat \sigma^+$ Fig. \ref{fig:spin1} shows logarithmic scaling of $S_2$ only for $\Delta=1$ because only then the total x- and y-magnetization are also conserved. Otherwise it indicates linear growth of $S_2$ with time. We can understand this now as a direct consequence of the Blombergen-de Gennes conjecture, which predicts a power law rather than an exponentially decaying ITAC in the isotropic case (see insets of Fig. \ref{fig:spin1}).
 We note, that
the regular and chaotic Hamiltonians used in the original work by Prosen and
\v{Z}nidari\v{c} \cite{Prosen2007a} have also been investigated numerically.
The OSRE shows the expected time dependence, i.e. logarithmic scaling in the
regular and linear scaling in the chaotic case. 

From the numerical results we can also extract the von Neumann entropy as a
function of time. It scales exactly as $S_2$ in the spin-$\frac12$ model for
all operators we looked at. The results are not conclusive in the spin-1
case however, since the dependence on the matrix dimension $\chi$ used in
the simulations is much stronger. At least they do not contradict the
presumption, that again the scaling is the same as for $S_2$.

In summary we have given a simple explanation of the at most logarithmic 
time dependence of the OSRE $S_2$ for the spin-1/2 XX model as a generic
integrable model that can be mapped to free fermions. The
operator dynamics in that model is equivalent to two uncoupled chains of
free fermions with an initial state corresponding to the operator under
consideration. For local operators these initial states are rather simple.
E.g. an operator $\hat \sigma^z_j$ corresponds to a single fermion in each
chain. We have shown that the bi-partite OSRE $S_2$ is strictly related to 
the fluctuations of the fermion number in the two partitions, which in turn allowed 
a simple understanding of the entropy dynamics. We have shown furthermore that
for any model, integrable or not, $S_2$ in OS can be bound by 
the infinite-temperature auto-correlation function of the considered operator.
This in turn means that for systems and observables for which the 
Blombergen-de Gennes conjecture of spin diffusion holds, an at most
logarithmic growth of the OS entanglement is expected.
The latter applies e.g. for local operators that constitute a global
conservation law.

We are indebted to Thoma\v{s} Prosen, Jesko Sirker, and Frank Verstraete for
valuable discussions, the SFB TRR49 of the DFG and the Excellence Initiative (DFG/GSC 266) for financial support, and the Erwin Schr\"{o}dinger Institute, Vienna, for hospitality.


\end{document}